\documentclass[sigconf,10pt]{acmart}
\usepackage[linesnumbered,lined,boxed,commentsnumbered,ruled,vlined]{algorithm2e}
\usepackage{graphicx}
\usepackage{booktabs}
\usepackage{multicol}
\usepackage{indentfirst}
\usepackage{url}
\usepackage{caption}
\usepackage{graphicx}
\usepackage{float} 
\usepackage{subcaption}
\graphicspath{{figs/}}

\begin{document}

\title{IMRSim: A Disk Simulator for Interlaced Magnetic Recording Technology}

% Fill in your details. The \documentclass "anonymous" parameter keeps them hidden, remove it after the review process finishes
\author{Zhimin Zeng}
\affiliation{%
  \institution{Huazhong University of Science and Technology}
  \city{Wuhan}
  \state{China}
}
\email{zengzhimin@hust.edu.cn}

\author{Xinyu Chen}
\affiliation{%
  \institution{Huazhong University of Science and Technology}
  \city{Wuhan}
  \state{China}
}
\email{cxy17340062501@163.com}

\author{Laurence T Yang}
\affiliation{%
  \institution{Huazhong University of Science and Technology}
  \city{Wuhan}
  \state{China}
}
\email{ltyang@hust.edu.cn}

\author{Jinhua Cui}
\affiliation{%
  \institution{Huazhong University of Science and Technology}
  \city{Wuhan}
  \state{China}
}
\email{csjhcui@gmail.com}

\begin{abstract}
\par The emerging interlaced magnetic recording (IMR) technology achieves a higher areal density for hard disk drive (HDD) over the conventional magnetic recording (CMR) technology. IMR-based HDD interlaces top tracks and bottom tracks, where each bottom track is overlapped with two neighboring top tracks. Thus, top tracks can be updated without restraint, whereas bottom tracks can be updated by the time-consuming read-modify-write (RMW) or other novel  update strategy. Therefore, the layout of the tracks between the IMR-based HDD and the CMR-based HDD is much different. Unfortunately, there has been no related disk simulator and product available to the public, which motivates us to develop an open-source IMR disk simulator to provide a platform for further research.

\par  We implement the first public IMR disk simulator, called IMRSim, as a block device driver in the Linux kernel, simulating the interlaced tracks and implementing many state-of-the-art data placement strategies. IMRSim is built on the actual CMR-based HDD to precisely simulate the I/O performance of IMR drives. While I/O operations in CMR-based HDD are easy to visualize, update strategy and multi-stage allocation strategy in IMR are inherently dynamic. Therefore, we further graphically demonstrate how IMRSim processes I/O requests in the visualization mode. We release IMRSim as an open-source IMR disk simulation tool and hope to attract more scholars into related research on IMR technology.
%In addition, IMRSim has very good scalability for data management in IMR.
\end{abstract}

\maketitle

\sloppy

\section{Introduction}
\par
\begin{figure}[ht]
    \centering
    \includegraphics[scale=0.45]{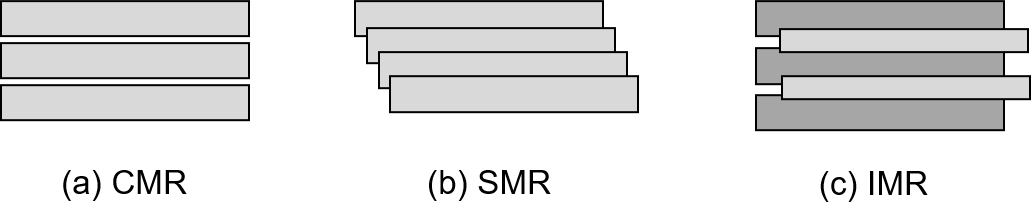}
    \caption{The different track layouts in HDDs.}
    \label{fig:tracklayout}
\end{figure}

 With the advent of the big data era, enterprises and industries need storage systems with larger capacity and lower costs. Affected by the superparamagnetic effect \cite{ref1}, the areal data density of CMR (see Fig. \ref{fig:tracklayout}(a)) has reached the limitation \cite{ref2,ref24}. To further expand disk storage capacity and reduce costs, academia and storage manufacturers are trying to explore new technologies and methodologies. Among the innovations in track layout, shingled magnetic recording (SMR) \cite{ref3,ref30,ref31} and interlaced magnetic recording (IMR) \cite{ref4,ref5,ref21} have gained tremendous popularity in disk devices \cite{ref29}. Furthermore, with the combination of the energy-assisted technologies for the recording head or media, such as heat-assisted magnetic recording (HAMR) \cite{ref6,ref7}, microwave-assisted magnetic recording (MAMR) \cite{ref8,ref9}, hard disk drives can further improve the areal data density.

\par In the SMR technology, to maximize the data storage density \cite{ref23}, the tracks overlap like roof tiles to shorten the track gap, as shown in figure \ref{fig:tracklayout}(b). As each track is partially overlapped with its subsequent track, random writes may corrupt data on the subsequent tracks. To protect the data, all subsequent tracks must be backed up before randomly writing a track, and then rewritten after randomly writing the track. This time-consuming phenomenon, known as read-modify-write (RMW), seriously degrades the random write performance of the SMR drive.

\par Compared to SMR, the alternative IMR technique (Fig. \ref{fig:tracklayout}(c)) achieves better random write performance. IMR organizes the tracks in an interlaced fashion rather than the tiled coverage, thus the random write process will only affect at most two adjacent tracks. As shown in Fig. \ref{fig:tracklayout}(c), IMR divides tracks into top and bottom tracks, where each bottom track is overlapped with two neighboring top tracks, and each top track partially covers two neighboring bottom tracks.

The essences of the IMR technique are energy-assisted recording and track overlapping. In IMR, the bottom track is wider than the top track \cite{ref4}. However, the traditional perpendicular magnetic recording (PMR) technology cannot control the width of the track. Thus, such track characteristics can only be achieved by the energy-assisted technology at present. The wider bottom track requires higher energy intensity than the narrower top track, which results in a higher data storage density in the bottom track than the top track \cite{ref10}. Therefore, data updating on one top track requires a lower energy intensity that is not enough to destroy the adjacent bottom tracks. As a result, IMR can perform update-in-place operation for top track data updating without the additional rewriting overhead. Unfortunately, higher energy intensity is required to update data on a bottom track. This will destroy data from the adjacent top tracks. To avoid the data corruption on the top tracks, the time-consuming update mechanism (e.g., read-modify-write) is required \cite{ref4}. Anyway, compared with SMR, the theoretical performance of IMR technology is outstanding. However, at present, the development of IMR-specific technology is still in the research stage, both in academia and industry. Meanwhile, there is no publicly available IMR simulator.

\par To solve this problem, we propose the first open-source IMR disk simulator, called IMRSim \cite{ref11}, as a block device driver in the Linux kernel. IMRSim effectively simulates the interlaced track layout, and supports many state-of-the-art data placement scheme, such as the two-stage allocation strategy, the three-phase allocation strategy. Furthermore, IMRSim provides an extensible user interface to flexibly adjust the device-specific parameters. In addition, IMRSim can process I/O requests in the visualization mode, greatly improving the understanding of the emerging IMR technology. Finally, We test IMRSim against CMR-based HDD on some real workloads to evaluate the I/O performance of the IMR-based HDD. The results show that IMR's update strategy brings significant performance loss which meet the expectation of the evaluation. \textit{To the best of our knowledge, this is the first public IMR disk simulator.} The main contributions are summarized as follows:

\begin{itemize}
    \item We implement the first public IMR disk simulator, IMRSim. IMRSim can effectively capture the key characteristics of IMR, including simulating interlaced tracks with different data densities, redirecting and sending bio requests to realistically simulate update strategies, etc. It should be noted that, since there are currently no IMR-based HDD products on the market, we have to ignore some physical factors that may affect performance (e.g., the performance impact of different write energy intensity).

    \item IMRSim is scalable. IMRSim provides good scalability for future related research. IMRSim provides an extensible user interface for adjusting the simulator parameters flexibly. And, users can modify the I/O path and configure their IMR optimization according to their needs.

    \item IMRSim is visual. IMRSim provides the visualization component for the logical data layout and the data movement in I/O requests, enabling users to gain a clearer understanding of the IMR technology. We hope visualization can expose new opportunities for future IMR optimization.
\end{itemize}

In the remaining of this paper, Section 2 presents the design details of IMRSim. Section 3 lists the experiment settings and analyzes the results. Section 4 presents more related work and Section 5 concludes this paper.

\section{IMR Simulation}
The design goal of the proposed IMR disk simulator, IMRSim, is to simulate an relatively accurate, scalable, and visual IMR-based HDD. To achieve this goal, we use the device-mapper (DM) framework to create a Linux kernel module, and then use the module to export a pseudo block device in user-space that acts like an IMR drive. As we build the device-mapper target directly on top of CMR-based HDD, the module will execute the incoming requests on the bottom CMR drive finally. Therefore, IMRSim has the same rotational speed and the head switch time as the CMR-based HDD \cite{ref22}. But, the logical arrangement of the tracks in IMRSim is different from the CMR-based HDD. IMRSim is a software application that simulates the interlaced track layout of an IMR-based HDD. Simultaneously, IMRSim provides a set of user interfaces, thus, users can verify status information and adjust simulator action. In addition, IMRSim provides the visualization component for the logical data layout and the
data movement in I/O requests, enabling users to gain a clearer understanding of the IMR technology.

\subsection{IMRSim Kernel Module}
\begin{figure}[ht]
    \centering
    \includegraphics[scale=0.45]{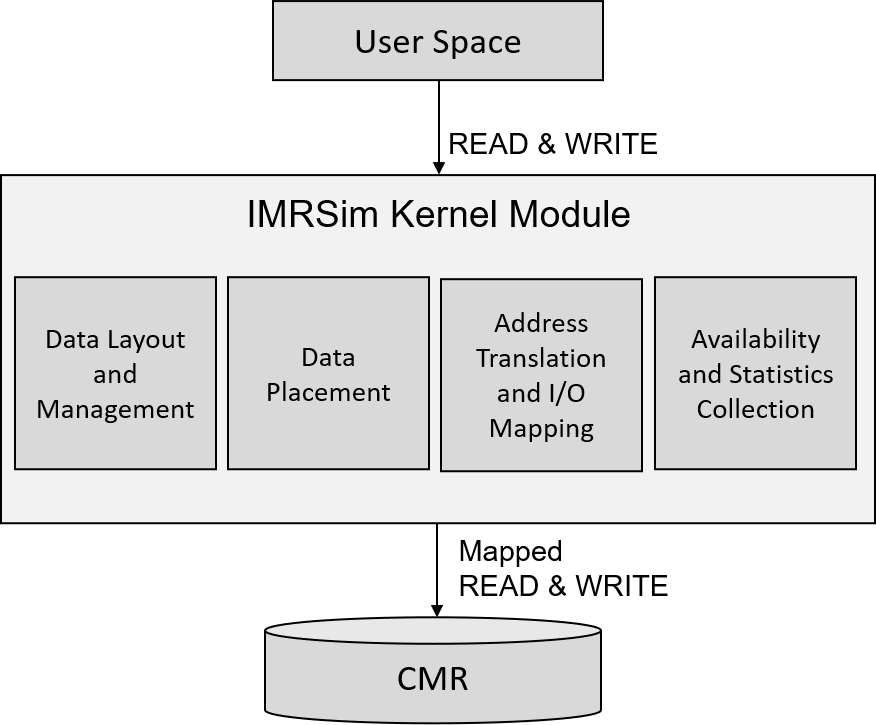}
    \caption{The architecture of IMRSim.}
    \label{fig:architecture}
    \vspace{-0.15in}
\end{figure}

Fig. \ref{fig:architecture} depicts the architectural overview of IMRSim. The read/write requests initiated by the upper-layer application are routed from top to bottom in Linux's I/O subsystem. The IMRSim kernel module has the ability to capture and modify the original request (e.g., redirect). Thus, a mapped request is obtained, and then the request is added to the request queue until it is processed on disk.

The IMRSim kernel module contains the following sub-modules: (1) the data layout and management sub-module, which simulates the layout of interlaced track and uses ``zone" to manage the track space (``zone" is similar to ``TG" \cite{ref25,ref28}), (2) the data placement sub-module, which handles different types of requests (read, write and update) for the simulated IMR disk, (3) the address translation and I/O mapping sub-module, which does the I/O redirection, (4) the availability and statistics collection sub-module, which launches a persistent thread to ensure availability,  synchronize meta-data, and collect status information of the simulator. We will introduce them in detail in the following.

\begin{figure}[ht]
    \centering
    \includegraphics[scale=0.4]{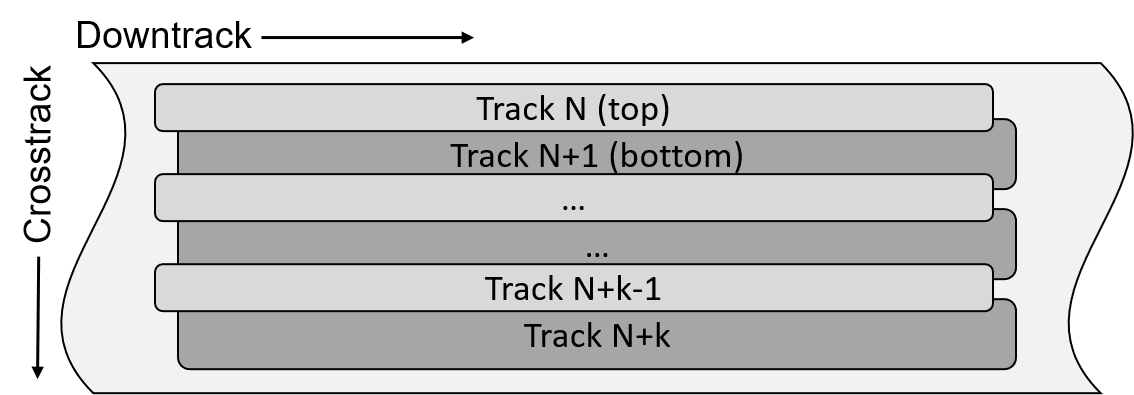}
    \caption{The data layout of zone.}
    \vspace{-0.15in}
    \label{fig:zone_layout}
\end{figure}

\noindent {\bf Data Layout and Management.}The data layout and management sub-module determines how the disk tracks and sectors are logically grouped together and managed. The data layout is organized in zones, which are clusters of several continuous interlaced tracks, as illustrated in Fig. \ref{fig:zone_layout}. IMR can consist of one or more zones. We assume that an application can be assigned to a zone and focus on data management within a zone. Note that the original intention of our design of zone is different from the design of band in SMR. Such a partition method can maintain the spatial locality of application data inside the zone to increase data access efficiency. Since IMRSim is a block device, a block (typically 4KB) is the smallest unit of management in the simulation. Each block is assigned a unique logical/physical block address.

\par 

\par It's worth noting that the bottom track is wider and has a larger data storage density than the top track. Thus, we design the top and bottom tracks with the different data densities. According to previous research \cite{ref7}, the data density of the bottom track in IMRSim is about 1.25 times that of the top track.

\noindent {\bf Data Placement.}
Data placement sub-module handles the incoming requests (e.g., read, write, or update). While the preceding discussion shows that, to avoid data loss or damage, before updating one bottom track data, IMR needs to protect the data in the affected top tracks. Recently, more and more different update strategies have emerged, such as read-swap-write (RSW) \cite{ref22} and move-on-modify (MOM) \cite{ref32}. IMRSim can easily extend these strategies, here for simplicity, IMRSim applies the traditional read-modify-write (RMW)  update strategy. 

\par The RMW process of data read, write and update operations is shown in Algorithm \ref{alg:algorithm1}. For the update of the data block, we need to judge the properties of the track where the block is located. If the block is on the top track, the update is free; if the block is on the bottom track, the RMW process is required during the update operation. It is worth noting that whether the top track data is backed up in memory or on disk has a greater impact on performance, but such analysis is beyond the scope of this article. In particular, in our implementation, we use page to keep data that needs to be backed up in memory.

\begin{algorithm}
\caption{The process of reading/writing.}
\label{alg:algorithm1}
\SetKwData{Lba}{lba}\SetKwData{Cdir}{cdir}\SetKwData{Bio}{bio}\SetKwData{WRITE}{WRITE}\SetKwData{READ}{READ}
\SetKwFunction{RMWStrategy}{RMWStrategy}\SetKwFunction{ReadBio}{ReadBio}\SetKwFunction{WriteBio}{WriteBio}
\SetKwInOut{Input}{input}
\Input{A bio structure $bio$.}
\BlankLine
\Lba $\leftarrow$ \Bio.bi\_sector\;
\Cdir $\leftarrow$ bio\_data\_dir(\Bio)\;
\uIf(){\Lba is illegal}{
\emph{report error}\;
}
\uIf{\Cdir == \WRITE}{
\uIf{\Lba is on bottom track and adjacent top blocks are valid}{
\RMWStrategy{\Bio}\;
}
\uElse{\WriteBio{\Bio}\;}
}
\uElse{\ReadBio{\Bio}\;}
\end{algorithm}\DecMargin{1em}

\noindent {\bf Address Translation and I/O Mapping.}
% Why IMR allocate not LBA=PBA? and Why Bottom first?
For traditional CMR, LBA (logical block address) is equal to PBA (physical block address) and it can write randomly. However, if the IMR is like the CMR, then there is the potential for severe write amplification problems in the low-utilization allocation phase. So, we let the IMR allocate from bottom tracks, and then allocate top tracks, so that no write amplification occurs during the allocation of the bottom track. In this way, IMR must needs to keep a mapping table (MT) to map the logical block address (LBA) to a physical block address (PBA). And this sub-module focuses on how to convert LBA to PBA.

% This sub-module is mainly responsible for the relevant conversion of the request address. This is because, IMR is specially designed to allocate the bottom tracks first, following by the top tracks \cite{ref5}, IMR must needs to keep a mapping table (MT) to map the logical block address (LBA) to a physical block address (PBA). In short, this sub-module focuses on two things: (1) how to convert LBA to PBA, and (2) how to record this mapping table.

%\par \begin{figure}[ht]
%    \centering
%    \includegraphics[scale=0.45]{address_translation.png}
%    \caption{ The address translation funciton.}
%    \label{fig:at_function}
%    \vspace{-0.15in}
%\end{figure}

%\par Here, we define two helpful functions as shown in Fig. \ref{fig:at_function}. 
\par Here, we define two helpful functions. The first one is the address translation function ($AT$, e.g., $y=AT(x)$) is responsible for converting the request address into the logical zone ID, the track offset and the block offset triples in the related zone. The second one is its inverse function (denoted as $AT^{-1}$) performs the reverse conversion. The $x$ represents LBA or PBA, and $y$ represents the relative block address, which is represented by a triple ($Zone\,ID$, $Track$ $Offset$, $Block\,Offset$).

% \par Then, according to zone information, the whole mapping table is divided into multiple sub-mapping tables. In each sub-mapping table, the block offset is used for mapping, so that the size of each record is limited to 0\textasciitilde65535 (since a zone has 65536 blocks). Also, for the future better scalability, we use int type to store a record. Therefore, a 4KB block requires 32 bits to record, and the space occupation of the mapping table is about 0.97\%. 

\par \begin{figure}[ht]
    \centering
    \includegraphics[scale=0.45]{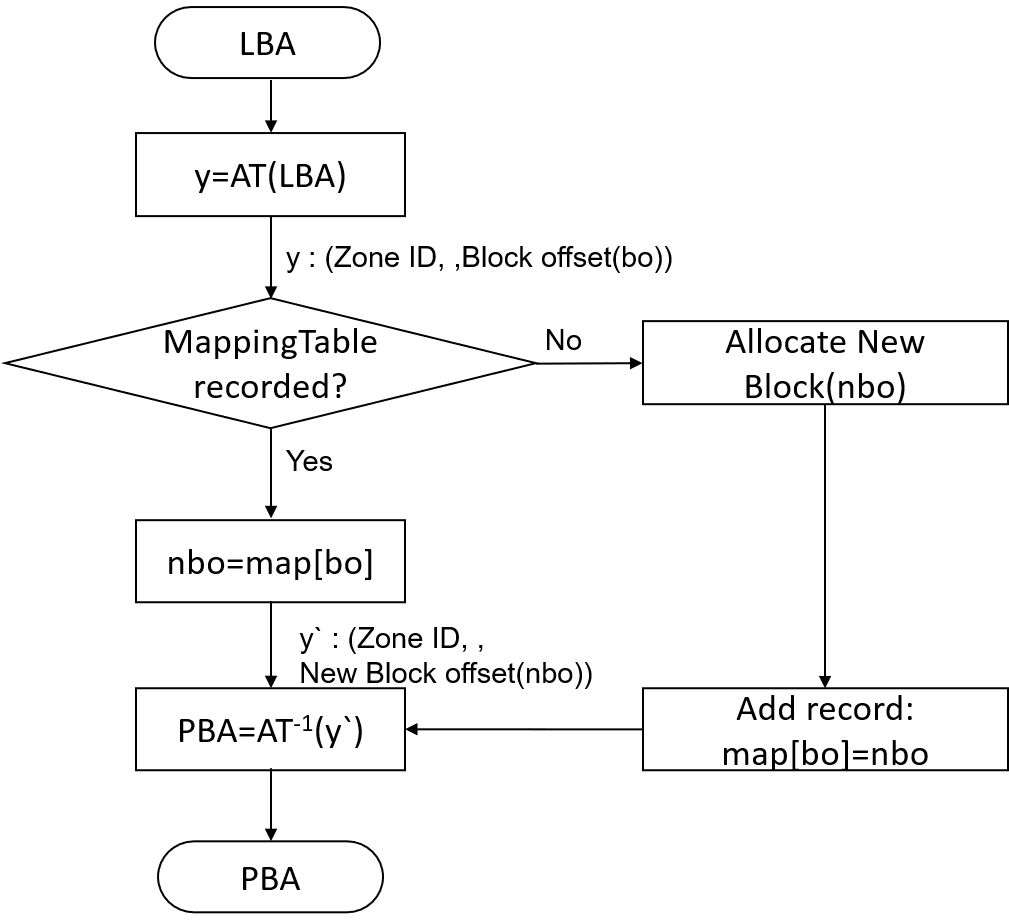}
    \caption{The I/O Mapping process.}
    \label{fig:io_mapping}
    \vspace{-0.15in}
\end{figure}

\par As illustrated in Fig. \ref{fig:io_mapping}, to get the PBA with the given LBA, IMRSim first uses $AT$ to convert the LBA into the managed triple to get the zone ID and block offset ($bo$). Then, there will be two cases:
% Note that the ``MappingTable" in Fig. \ref{fig:io_mapping} refers to the sub-mapping table corresponding to $Zone\,ID$. 

\par 1) If there is no record corresponding to $bo$ in the MT, it means a new write operation. At this time, IMRSim allocates a new block (with a new block offset, i.e. $nbo$)  according to the different track allocation schemes, and then record $bo$ (block offset) to $nbo$ in the MT.
\par 2) If there is a record corresponding to $bo$ in the MT, it means an update operation. Thus, the $nbo$ can be obtained by querying the MT. 
\par Finally, the PBA can be obtained by using $AT^{-1}$.

Note that the address translation triples (ie, y and y\`) omit the track offset in Fig. \ref{fig:io_mapping}. This is because, with the $zone ID$ and $nbo$ can uniquely identify the PBA. Currently, we have implemented the classical two phases allocation strategy, and the three phases allocation strategy in IMRSim. Of course, the other multi-phase allocation strategies can be designed as needed. Also, our mapping table may be utilized to implement strategies such as track caching, hot and cold data swap, and track flipping \cite{ref27,ref28}.

\noindent {\bf Availability and Statistics Collection.}
This sub-module provides a certain level of availability for disks, and collects some simulator status information.
\par We add a meta-data area after data storage to record the mapping table and disk statistics. In addition, users can easily add new statistical indicators in IMRSim. In the current version of IMRSim, we record the behavior information of the simulator (e.g., the number of writes that occurred additionally) into memory, and open a persistent thread to periodically flush the collected statistics or the changed metadata to the disk.

\subsection{User Interface}
\par To facilitate user interaction with the simulator, we design a command-line user interface. It is a standard C program implemented based on the ioctl system call. The ioctl system call can be easily extended in the kernel, which makes IMRSim highly extensible. The user can use such system calls to flexibly adjust the design parameters and control the behavior of the simulator according to their needs. For example, users can input ``./imrsim\_util /dev/mapper/imrsim l 5 3'' to set the allocation strategy to three-phase allocation.

\par In the implementation, the interface needs some command parameters to select the required operation. If the input format is wrong, it will notify the user of the wrong format and provide a friendly help interface and a correct input example.

\begin{figure}[h!]
    \centering
    \begin{subfigure}[h]{0.23\textwidth}
           \centering
           \includegraphics[width=\textwidth]{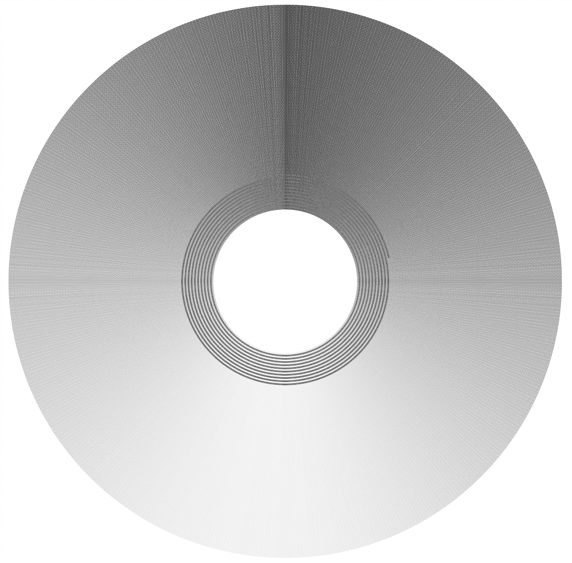}
            \caption{Allocate blocks.}
            \label{fig: visual1}
    \end{subfigure}
    \begin{subfigure}[h]{0.23\textwidth}
            \centering
            \includegraphics[width=\textwidth]{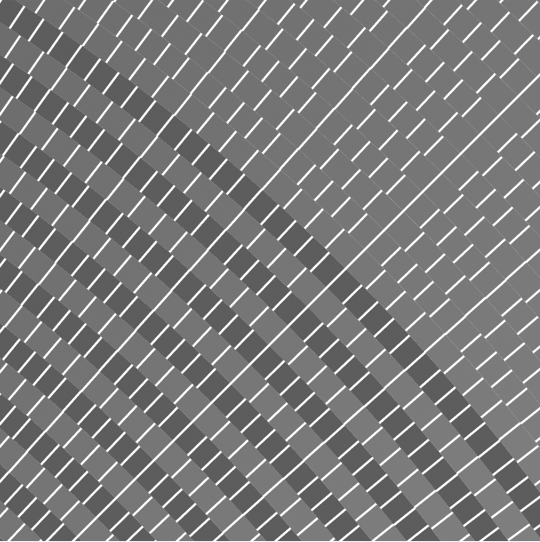}
            \caption{Detail part of the zone.}
            \label{fig: visual2}
    \end{subfigure}
    \caption{Handing write requests when two-stage IMR is empty.}
	\label{fig: visualization} 
\end{figure}

\subsection{Visualization}
In addition, we implement a quick visualization of IMRSim which cam briefly demonstrate the dynamic process of IMR processing requests. Such visualization can help readers to further understand the track characteristics of IMR. We have collected the I/O traces of a zone during the fio test to realize visualization. As shown in Fig. \ref{fig: visual1}, a two-stage IMR zone always allocates the bottom track before the top track when it is empty to process write requests (whether sequential or random). There is always one track spaced apart during the allocation process (the allocated track turns darkgray), as shown in Fig. \ref{fig: visual2}.

\section{Simulation Analysis}
\subsection{Experimental Methodology}
\par This section evaluates the performance of two different allocation strategies. We created a 128GB IMR-based HDD and a 128GB CMR-based HDD on a Western Digital's pure CMR drive (see Table \ref{tab:tested_device}). Note that we use the same partition on the CMR drive to ignore the impact of OD/ID differences on performance. In our experiment, we implement different allocation strategies for evaluation as follows: First, we evaluate the performance of IMR two-stage allocation versus three-stage allocation strategies. Second, the typical RMW update strategy is implemented.

\begin{table}[h!]
    \centering
    \caption{The information for tested device.}
    \label{tab:tested_device}
    \begin{tabular}{c c}
        \toprule
        \multicolumn{2}{c}{HDD Details} \\
        \midrule
        HDD model & WD20EJRx \\
        drive cache size & 64MB \\
        rotational speed & 5400rpm \\
        \toprule
        \multicolumn{2}{c}{IMRSim Parameters} \\
        \midrule
        block size & 4096B \\
        top track size & 456 blocks \\
        bottom track size & 568 blocks \\
        capacity & 128GB \\
        \bottomrule
    \end{tabular}
\end{table}

\begin{table}[h!]
    \caption{The statistics of the evaluated workloads.}
    \label{tab:stats_traces}
    \resizebox{\linewidth}{!}{
    \begin{tabular}{|c|c|c|c|}
    \hline
    workload & Number of Read / Write Requests & Total Read / Write Size (GB) & write ratio \\ \hline
    hm\_0 & 1,417,748 / 2,575,568 & 9.96 / 20.48 & 64.50\% \\ \hline
    proj\_0 & 527,381 / 3,697,143 & 8.97 / 144.27 & 87.52\% \\ \hline
    src1\_2 & 484,079 / 1,423,694 & 8.82 / 44.14 & 74.63\% \\ \hline
    src2\_2 & 350,930 / 805,955 & 22.79 / 32.28 & 69.67\% \\ \hline
    \end{tabular}}
\end{table}

\par We conduct all the experiments on a desktop PC, which is equipped with twelve Intel(R) Core(TM) i5-10400 CPU @ 2.90GHz processors and 16 GB DDR4 DIMM memory, where the operating system is 64-bit Ubuntu 14.10 with Linux kernel 3.16.0. For performance testing, we use fio-3.30 to replay and evaluate four real write-intensive workloads, i.e., hm\_0, proj\_0, src1\_2 and src2\_2, collected by Microsoft Research Cambridge \cite{ref32}. Table \ref{tab:stats_traces} provides more details about these workloads including the numbers of read/write requests, the total read/write sizes (GB), and the write ratio. In particular, to better evaluate the performance in handling requests, four write-intensive workloads with the accessed LBAs fitting in the tested disk space (i.e., 128GB).
\par In the experimental configuration, the number of threads is set to 1, the IO queue depth is set to 32, and the IO engine is set to libaio. In addition, to avoid the impact of kernel cache on the simulator performance, all the I/O requests generated by the test are the direct I/O, by passing the kernel buffer. Moreover, we execute command {\textcolor{gray}{hdparm -W0 -A0 -a0}} to have the HDD cache disabled. One more thing to note, since there are no rewrites operation are incurred if the space usage is less than 55.47\% of the disk space (at this point, the first stage has just been allocated). Thus, we use 32 KB (which is close to the average write size of the four evaluated workloads) random write requests to initialized the simulated IMR-based HDD. 

\subsection{Experimental Results}

\begin{figure}[b]
    \centering
    \includegraphics[scale=0.65]{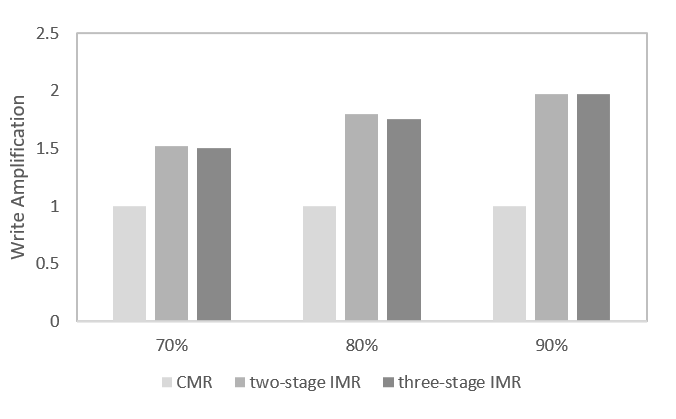}
    \caption{Write amplification under the src1\_2 workload.}
    \label{fig:wa_usage}
\end{figure}

\begin{figure*}[t!]
    \begin{subfigure}[t]{0.3\textwidth}
           \centering
           \includegraphics[width=\textwidth]{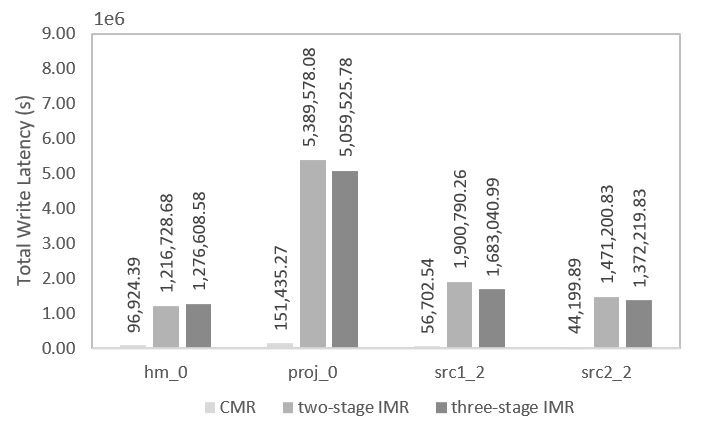}
            \caption{Total Write Latency (80\% Usage)}
            \label{fig: perf_write_latency}
    \end{subfigure}
    \begin{subfigure}[t]{0.3\textwidth}
            \centering
            \includegraphics[width=\textwidth]{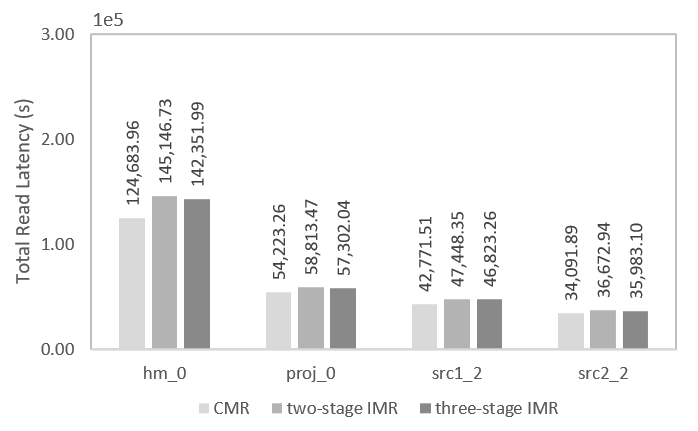}
            \caption{Total Read Latency (80\% Usage)}
            \label{fig: perf_read_latency}
    \end{subfigure}
    \begin{subfigure}[t]{0.3\textwidth}
            \centering
            \includegraphics[width=\textwidth]{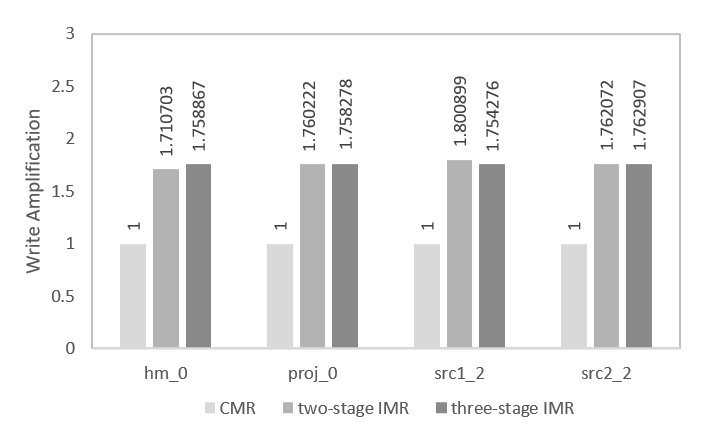}
            \caption{Write Amplification (80\% Usage)}
            \label{fig: perf_wa}
    \end{subfigure}
    \caption{Performance results with different workloads under 80\% space usage.}
	\label{fig: performance_traces} 
\end{figure*}

\par 1) \emph{Write Amplification under Different Space Usages:} Fig. \ref{fig:wa_usage} shows the write amplification factor with different space usage under the src1\_2 workload, where the x-axis denotes space usage of different tested devices and the y-axis represents write amplification factor (i.e., the ratio of the actual total number of writes to the number of write requests). Since the CMR-based HDD (denoted as \textcolor{gray}{CMR}) uses a cost-free in-place update strategy and does not cause write amplification problems, its write amplification factor is always 1. We use our IMRSim to simulate IMR-based HDD with two stage allocation strategy (denoted as \textcolor{gray}{two-stage IMR}) and IMR-based HDD with two stage allocation strategy (denoted as \textcolor{gray}{three-stage IMR}). 
\par It can be clearly observed that, as the space utilization increases, the write amplification factor gradually increases. Also, the write amplification of \textcolor{gray}{two-stage IMR} is slight larger than that of \textcolor{gray}{three-stage IMR}. However, as the space utilization increases, the write amplification difference between the \textcolor{gray}{two-stage IMR} and \textcolor{gray}{three-stage IMR} is getting smaller.

\par 2) \emph{Performance results under different workloads:} As shown in Fig. \ref{fig: performance_traces}, We evaluate total write and read latency and write amplification with different workloads under 80\% space usage. Specifically, we reply four real write-intensive workloads (i.e., hm\_0, proj\_0, src1\_2 and src2\_2) to collect the experimental results and verify the correctness of our simulator. 

\par {\bf Write Perfomance.} Fig. \ref{fig: perf_write_latency} demonstrates the write performance of three simulated devices under 80\% space usage, where the x-axis denotes various workloads and the y-axis represents the write performance in terms of the total write latency (i.e., the total time for executing write requests). 
\par It can be firstly observed that, under four emulated workloads, the simulated IMR-based HDD have a relatively large write performance gap with \textcolor{gray}{CMR} while the \textcolor{gray}{two-stage IMR} and \textcolor{gray}{three-stage IMR} achieve similar write performance. Specifically, \textcolor{gray}{CMR} outperforms IMR by 12.5x in the hm\_0 workload, by 35.6x in the proj\_0 workload, and by an average of 28.7x in four workloads. For the comparison of IMR-based HDDs with different allocation stages, \textcolor{gray}{two-stage IMR} outperforms \textcolor{gray}{three-stage IMR} in write performance in the hm\_0 workload, however it performs worse compared to \textcolor{gray}{three-stage IMR} in the other three workloads and has an average write performance loss of 0.55x in four workloads.
\par The experimental result shows that the performance loss brought by the RMW update strategy is very huge, although each update behavior only causes at most two additional read and write operations. In order to analyze the reasons for the huge performance loss caused by RMW, we calculated the write amplification factor of each workload under 80\% space usage in the experiment. We count the number of extra writes in blocks and calculate the write amplification factor, and the result is shown in Fig. \ref{fig: perf_wa}. It can be observed that, across four emulated workloads, the additional writes of IMR are at most 1.8x that of \textcolor{gray}{CMR}. A mere 1.8x difference in write times results in a 28x difference in write performance. This is mainly because mechanical disks consume time in seek and rotation delays, and the RMW process needs to spend extra time on these positioning operations. The extra time is much larger than the data transfer delay. For IMR-based HDDs with different allocation stages, the average extra write times of the \textcolor{gray}{two-stage IMR} is slightly larger than that of the \textcolor{gray}{three-stage IMR}, which results in a slightly worse write performance than the \textcolor{gray}{three-stage IMR}.

\par {\bf Read Perfomance.} Fig. \ref{fig: perf_read_latency} demonstrates the read performance of three simulated devices under 80\% space usage, where the x-axis denotes various workloads and the y-axis represents the read performance in terms of the total read latency (i.e., the total time for executing read requests). 
\par It can be clearly observed that, in four emulated workloads, \textcolor{gray}{two-stage IMR} and \textcolor{gray}{three-stage IMR} exhibit read performance close to \textcolor{gray}{CMR}, as expected. Specifically, the read performance of \textcolor{gray}{two-stage IMR} is 9.55\% slower than that of \textcolor{gray}{CMR} on average under 80\% space usage, and the read performance of \textcolor{gray}{three-stage IMR} is only 7.98\% slower than that of \textcolor{gray}{CMR} on average. The main reason for the slight difference in read performance between IMR and \textcolor{gray}{CMR} is that IMR needs to maintain a dynamic mapping table, and different allocation methods will also lead to different block addresses, which in turn affects delays such as seek time.

\section{Related Works}
\par We have collected some studies related to disk simulators. DiskSim \cite{ref16} is a disk simulator developed by Carnegie Mellon University. It can accurately simulate the performance of various types of traditional hard disks, but it is only suitable for running in 32-bit systems. Although 64-bit systems are supported later, it is complicated to use. To support solid state drives (SSD), Microsoft created an SSDmodel module \cite{ref17} to simulate the performance of SSD, and it has been widely used. FlashSim \cite{ref18} is another emulator that evaluates SSD performance and studies some flash translation layers used in SSDs. Tan et al .\cite{ref19} designed the SMR simulation platform for various testing and analysis of shingled translation layer designs. Pitchumani et al. \cite{ref20} developed a tool to imitate SMR disk by implementing a shingled device-mapper target on top of the Linux kernel's block devices. Their goal is to allow STL to be evaluated on actual hard drives. So far, there is no simulation tool that is publicly available to simulate IMR disk and analyze their performance with emerging allocation and update strategies.

\section{Conclusion}
\par In this paper, we implement the first open-source IMR disk simulator, IMRSim, to simulate several different interlaced track layouts and accurate I/O performance. The performance of IMR is excellent in theory, but in practice, the time-consuming RMW process costs more than we expected. Further, IMRSim is designed in Linux kernel space using the Device Mapper framework, and IMRSim exposes a scalable user interface for interaction between the user and the disk simulator. Finally, we provide a visualization component to interestingly demonstrate how IMR handles I/O requests.

\bibliographystyle{ACM-Reference-Format}
\bibliography{references}
\end{document}